\documentclass[preprint,12pt]{elsarticle}

\usepackage{amssymb}
\usepackage{mdsymbol}
\usepackage{multirow}
\usepackage{siunitx}
\usepackage{amsmath}

\journal{Pattern Recognition}

\begin{document}

\begin{frontmatter}



\title{GPU-Accelerated RSF Level Set Evolution for Large-Scale Microvascular Segmentation}


\author[inst1]{Meher Niger}
\author[inst1]{Helya Goharbavang}
\author[inst2]{Taeyong Ahn}
\author[inst2]{Emily K. Alley}
\author[inst2,inst3,inst4,inst5,inst6]{Joshua D. Wythe}
\author[inst1]{Guoning Chen}
\author[inst1]{David Mayerich}

\affiliation[inst1]{organization={Electrical and Computer Engineering, University of Houston},%Department and Organization
            addressline={ 4226 Martin Luther King Boulevard}, 
            city={Houston},
            postcode={77204}, 
            state={TX},
            country={United States}}

\affiliation[inst2]{organization={Department of Cell Biology, University of Virginia School of Medicine},%Department and Organization
            addressline={Charlottesville}, 
            city={VA},
            {USA}}
\affiliation[inst3]{organization={Department of Neuroscience, University of Virginia School of Medicine},%Department and Organization
            addressline={Charlottesville}, 
            city={VA},
            {USA}}
\affiliation[inst4]{organization={Robert M. Berne Cardiovascular Research Center, University of Virginia School of Medicine},%Department and Organization
            addressline={Charlottesville}, 
            city={VA},
            {USA}}
\affiliation[inst5]{organization={Brain, Immunology, and Glia (BIG) Center, University of Virginia School of Medicine},%Department and Organization
            addressline={Charlottesville}, 
            city={VA},
            {USA}}
\affiliation[inst6]{organization={UVA Comprehensive Cancer Center, University of Virginia School of Medicine},%Department and Organization
            addressline={Charlottesville}, 
            city={VA},
            {USA}}
\begin{abstract}
Microvascular networks are challenging to model because these structures are currently near the diffraction limit for most advanced three-dimensional imaging modalities, including confocal and light sheet microscopy. This makes semantic segmentation difficult, because individual components of these networks fluctuate within the confines of individual pixels. Level set methods are ideally suited to solve this problem by providing surface and topological constraints on the resulting model, however these active contour techniques are extremely time intensive and impractical for terabyte-scale images. We propose a reformulation and implementation of the region-scalable fitting (RSF) level set model that makes it amenable to three-dimensional evaluation using both single-instruction multiple data (SIMD) and single-program multiple-data (SPMD) parallel processing. This enables evaluation of the level set equation on independent regions of the data set using graphics processing units (GPUs), making large-scale segmentation of high-resolution networks practical and inexpensive. 

We tested this 3D parallel RSF approach on multiple data sets acquired using state-of-the-art imaging techniques to acquire microvascular data, including micro-CT, light sheet fluorescence microscopy (LSFM) and milling microscopy. To assess the performance and accuracy of the RSF model, we conducted a Monte-Carlo-based validation technique to compare results to other segmentation methods. We also provide a rigorous profiling to show the gains in processing speed leveraging parallel hardware. This study showcases the practical application of the RSF model, emphasizing its utility in the challenging domain of segmenting large-scale high-topology network structures with a particular focus on building microvascular models.
\end{abstract}

\begin{graphicalabstract}
\includegraphics[width=\textwidth]{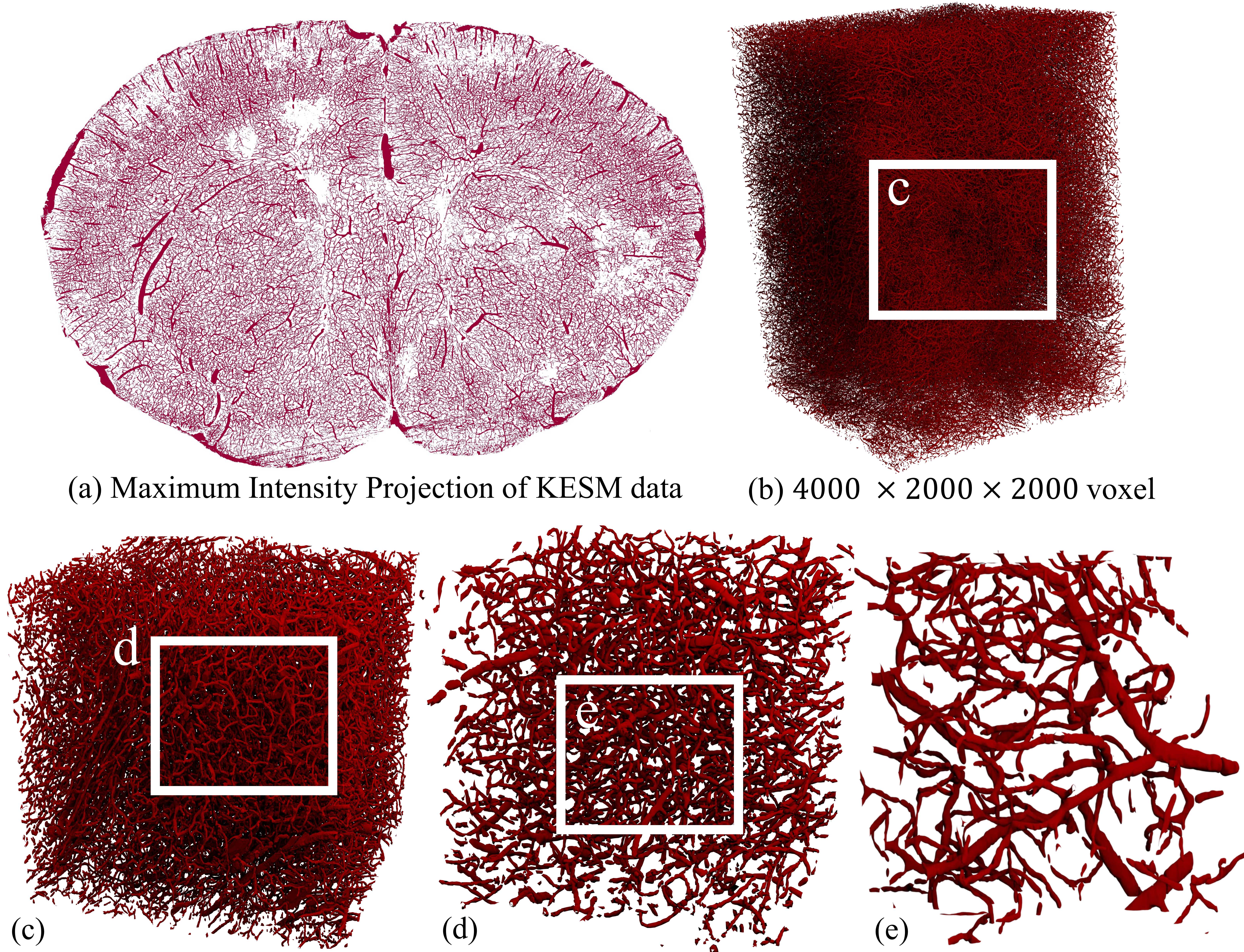}
\end{graphicalabstract}

\begin{highlights}
\item Reformulation of the Region-Scalable Fitting (RSF) level set model for SIMD and SPMD parallelism
\item Demonstration segmenting images of microvascular networks that are extremely large and topologically complex
\end{highlights}


\end{frontmatter}


%% main text\section{Introduction}
Microvascular networks play a critical role in tissue regulation and disease progression, however their three-dimensional microstructure is poorly understood. This is largely due to limitations in traditional 3D imaging methods, which lack the resolution to fully reconstruct vessels smaller than \SI{10}{\micro\meter} in diameter for large tissue blocks. As advances in light-sheet \cite{hsu2022ez} and milling microscopy \cite{knott2008serial} improve resolution and acquisition rates, the bottleneck is increasingly shifting to a lack of robust algorithms to build network models from these images.  

Most vascular segmentation techniques are designed for large vessels imaged \textit{in vivo} using retinal fundus imaging, magnetic resonance imaging (MRI), and micro-computed tamography (micro-CT) \cite{heller2021state,hong2020development}. Open source software for studying microvessel morphology, such as VesselVio \cite{bumgarner2022open} and TubeMap \cite{renier2016mapping,kirst2020mapping}, rely on a segmentation pipeline that is challenging to optimize because microvessel diameters ($\approx$\SIrange{3}{10}{\micro\meter}) are near the current resolution limits of most 3D fluorescent microscopes. Individual microvessels also form complex interconnected networks that span whole organs \cite{mayerich2022computational}. The structural complexity, combined with these resolution requirements, makes segmentation algorithms highly sensitive to noise and other artifacts.

Filtering methods such as the Hessian-based \textit{vesselness} filter have been proposed \cite{frangi1998multiscale} and continuously improved \cite{mahapatra2022novel} to enhance vessel contrast. However even minor changes in intensity and contrast across a large volume have a significant impact on segmentation accuracy. Semantic segmentation leveraging deep neural networks to enhance contrast have also been explored \cite{saadatifard2020cellular}, however these are difficult to generalize and fail to accurately reconstruct the vessel surface at sub-pixel accuracy. However, applying pre-trained filters (such as convolutional neural networks) to microvascular images may provide effective pre-processing for the proposed active contour approach.

Active contour models such as level sets \cite{sethian1999level} can take advantage of known structural features, such as interconnectivity and the shape of vessel cross-sections. Region-based force terms \cite{li2008minimization} are extremely robust in noisy and low-contrast data, providing the potential for a more consistent and deterministic segmentation.

However, these methods are computationally intensive and come with certain limitations. The vast majority focus on two-dimensional segmentation. While three-dimensional approaches have been studied in recent years \cite{chen2019distance,liu2019weighted,shu2021neighbor,wang2021review}, they are only demonstrated on very low-resolution images acquired using MRI and micro-CT. 

To address the computational challenges of existing level set models, we provide a formulation of the state-of-the-art region scalable fitting (RSF) model for three dimensions that enables highly-parallel execution. We also develop a seed surface method for initialization and demonstrate the scalability and accuracy of this approach on a 16GB microvascular image acquired using knife-edge scanning microscopy (KESM)
\cite{mayerich2008knife,mayerich2011fast}.

\section{Level Set Method}
Level set methods are a mathematical framework for tracking contours across uniform grids, such as two- and three-dimensional images. The distance to a contour is stored at each grid point $\phi(\mathbf{x})$ \cite{osher1988fronts}. The contour evolves over time by calculating the partial derivative $\frac{\delta\phi}{\delta t}$ and using it to update $\phi$. At any time $t$, the embedded surface is reconstructed by finding the isocontour $\Omega$ where $\Omega \in \phi(\mathbf{x})=0$. The level set function $\phi$ is usually represented using a signed distance function, where $\phi(\mathbf{x}) < 0$ inside of $\Omega$, which enforces topological consistency and facilitates finding the isocontour $\Omega$. 

The contour is moved over discrete time steps $t_n$ by calculating the gradient with respect to time and updating $\phi$:
\begin{equation}
\phi_{n+1} = \phi_n + \Delta_t E(\phi, \cdots)
\end{equation}
where
\begin{itemize}
    \item $\phi_n$ is the level set function at the current time step $n$
    \item $\phi_{n+1}$ is the updated level set
    \item $\Delta_t$ is the time step
    \item $E(\phi, \cdots)$ is an energy function defining how the contour changes over time
\end{itemize}

The update function $E$ can be thought of as the partial derivative of $\phi$ with respect to $t$. A level set method calculates $E$ to accomplish a desired goal, such as image segmentation. Common components of $E$ include regularization, smoothing, and force terms:
\begin{equation}
\label{eqn:levelset}
    E(\phi, \cdots) = R(\phi) + \alpha S(\phi) + \beta F(\cdots)
\end{equation}
where $\alpha$, and $\beta$ are constants defining the proportional contribution of each term.

\subsection{Regularization}
The regularization term ensures that $\phi$ maintains its characteristic as a signed distance function. Without regularization, level set functions exhibit rapid growth on both sides of the contour \cite{li2005level}, leading to unexpected movement and discontinuities. Early methods to address this problem rely on periodic reinitialization \cite{osher2005level} by calculating a new signed distance function using Fast Sweeping \cite{zhao2005fast} or Fast Marching \cite{sethian1999fast}. However, re-initialization is time-consuming and therefore only used periodically, leading to discontinuities in $\phi$ over time.

Regularization terms are integrated into the energy function and penalize deviations from $|\nabla\phi|=1$ to maintain a signed distance field \cite{li2010distance}. Li et al. \cite{li2010distance} proposed a regularization term to avoid the reinitialization of the level set function, while Wu et al. \cite{wu2012improved} introduced an extra regularization term for the edge-based level set method.

We employed forward and backward (FAB) diffusion $\phi$ \cite{li2008minimization}, which has historically good performance for region-based evolution \cite{chen2019distance,wang2014enhanced}.
\begin{equation}
R(\phi) = \nabla^2\phi-\frac{\nabla^2\phi}{|{\nabla\phi}|}
\end{equation}

\subsection{Smoothing}
The smoothing term creates an \textit{elastic} effect that minimizes the length or surface area of the contour $\Omega$ by penalizing regions of high curvature \cite{li2008minimization}: 
\begin{equation}
S(\phi) = \text{}\frac{\nabla^2\phi}{|{\nabla\phi}|}
\end{equation}
This term provides one of the major advantages of level set segmentation over other semantic approaches by fitting a smooth sub-diffraction surface to the image.

\subsection{Force}
The final term defines a force applied to the contour to achieve a desired goal such as image segmentation. Force terms fall into two categories: (1) edge-based and (2) region-based. Edge-based approaches apply a force to the contour based on its proximity to an edge. An \textit{edge stop function} is often used to slow the contour in the presence of a local edge. This can be approximated using the local gradient \cite{szeliski2022computer}, edge detection (Insight Toolkit) \cite{yoo2004insight}, or statistically-based methods that are more robust to noise \cite{liu2019weighted}. 

More recent work suggests that region-based approaches exhibit better performance \cite{wang2021review} in segmenting objects with complex boundaries. The earliest region-based approaches include work by Chan and Vese \cite{chan2001active} that minimize an integral across the interior and exterior of $\Omega$. However, this model faces challenges when the internal or external intensities are not homogeneous.

%While efforts to segment inhomogeneous structures have been explored \cite{paragios2002geodesic,vese2002multiphase}, for not containing any local intensity information as well.

Region Scalable Fitting (RSF) \cite{li2008minimization} was proposed to overcome these challenges by integrating across two fitting functions that regionally estimate pixel values on either side of $\Omega$. These fitting functions capture the local mean intensity values adjacent to the contour. More recent advances attempt to capture highly heterogeneous structures using clustering \cite{min2016novel} or statistically-based (Gaussian) approaches \cite{yu2018novel}. The current state-of-the-art, Local Approximation of Taylor Expansion (LATE) \cite{min2018late}, extends RSF using a first-order polynomial approximation to the interior intensity.

However, internal heterogeneities are not a major challenge for microvascular models, which have a high surface-to-volume ratio.  We therefore build directly on the more fundamental RSF model to take advantage of speed. We provide a detailed explanation of RSF along with a novel generalization to higher dimensions in Section \ref{sec:rsf}.

\subsection{Large-Scale Data}

Yang et al. proposed a 3D level set method based on RSF to segment low-topology structures in low-resolution brain MR images. Their approach relies on multiple registered atlases to improve the force term \cite{yang2019multi}. A reaction-diffusion level set method was proposed by Zhang et al. \cite{zhang2020reaction} for three-dimensional segmentation on computed tomography (CT) images. These data sets also segment low-topology structures in relatively low resolution (377 $\times$ 297 $\times$ 306) images with few artifacts. Another approach using a split Bregman method for medical image segmentation was proposed by Shu et al. \cite{shu2021neighbor} for relatively small (160 $\times$ 160 $\times$ 10) 3D cardiac MR images. Finally, Zhang et al. introduced a deep learning multi-atlas registration method \cite{zhang2021deep} to achieve a Dice score of $82\%$. In addition to the required CNN training, the technique also relies on multiple atlases that are unavailable, and likely impossible, for data at cellular resolution.

One of the major challenges with high-resolution 3D microscopy data is the volume size. In addition to parallel segmentation, level set methods may also be amenable to GPU acceleration. A model incorporating YOLOv4 and a region-based active contour model has been proposed by Dlamini et al. \cite{dlamini2023complete} for segmenting tumors in 260 $\times$ 260 $\times$ 30 lung cancer MR images with the Chan and Vese \cite{chan2001active} and Mumford Shah \cite{mumford1989optimal} models. While implementation details were not provided in the manuscript, these algorithms may be readily parallelized. However, they are sub-optimal when compared to more recent region-based approaches \cite{wang2021review}.

In this paper, we propose a parallel GPU-based implementation of region-based level sets using RSF as a foundation. We extend RSF to multiple dimensions and describe our approach for an efficient GPU-based implementation. In addition, we provide techniques for seeding and segmentation of gigavoxel 3D microscopy images, including profiling and validation of the segmented results.

% \david{We use the region scalable fitting model as the basis for our work because...(justify its use in the last paragraph here)}
\section{3D Region-Scalable Fitting (RSF)}
\label{sec:rsf}
Our proposed approach relies heavily on the RSF model introduced by Li et al.  \cite{li2008minimization}. Our modifications to the theory include (1) extension to higher dimensions ($D>2$) and (2) a formulation to enable SIMD and SPMD parallelism.

The region-based approach for image segmentation proposed by Chan and Vese \cite{chan2001active} calculates the force by minimizing the cost function:
\begin{equation}
    C(\Omega, c_-, c_+)=\oiiint_{\Omega^-}\left|I(\mathbf{x}) - c_- \right|^2 d\mathbf{x} + \oiiint_{\Omega^+}\left|I(\mathbf{x}) - c_+ \right|^2 d\mathbf{x}
\end{equation}
across the two regions inside ($\Omega^-$) and outside ($\Omega^+$) of the contour $\Omega$. In this formulation, $I$ is the input image and $c_-$ and $c_+$ are the average intensities inside and outside of the contour, respectively.

\begin{figure}[h]
    \centering   
    \includegraphics[width=0.75\textwidth]{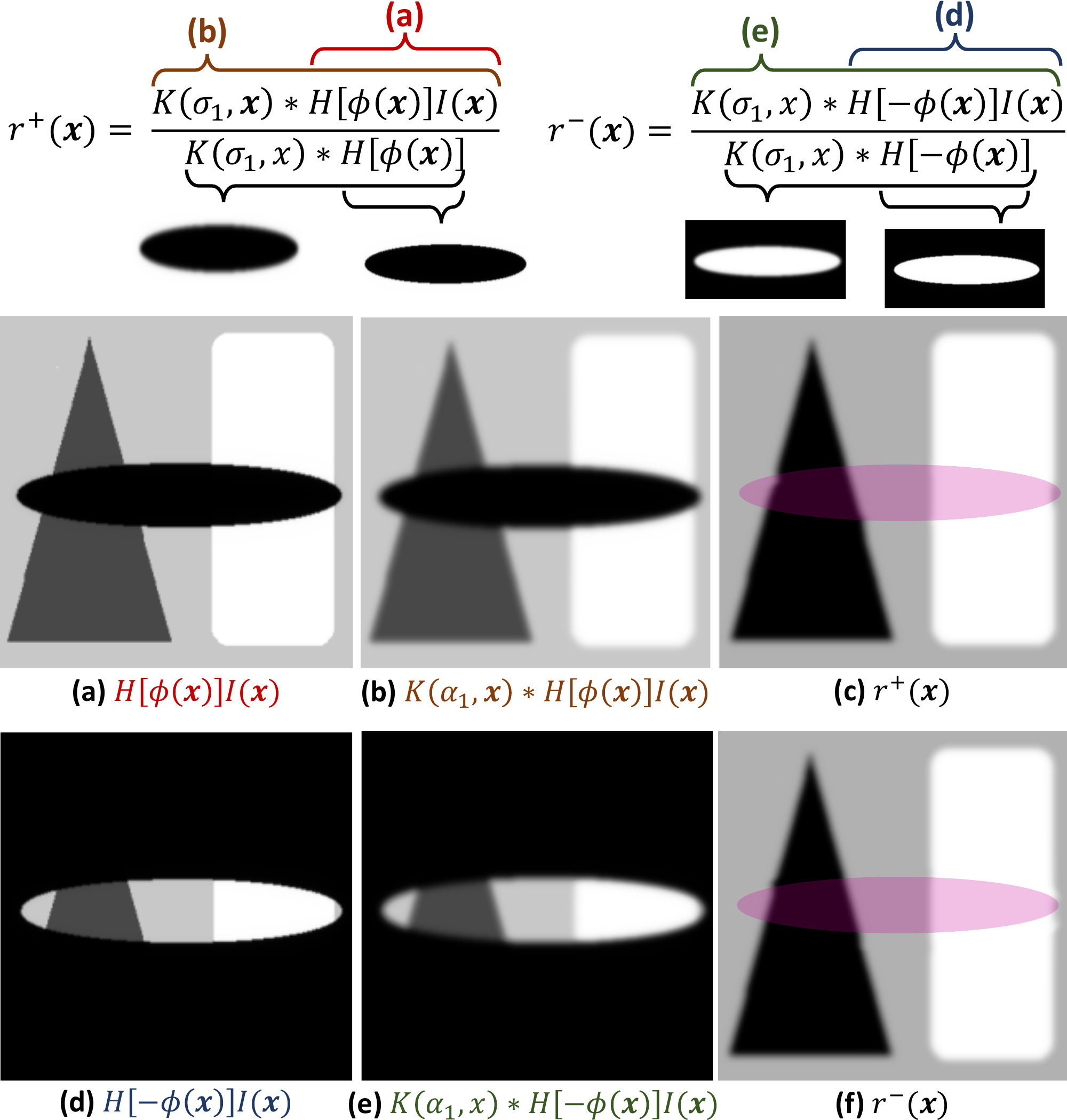}
    \caption{Calculation of the region-based intensities $r^+$ and $r^-$. (a, d) The input image is multiplied with the appropriate Heaviside function, and then (b, e) convolved with a Gaussian kernel. (c, f) The final images approximate the local intensity of $\Omega$ outside (top) and inside (bottom) of the contour.}
    \label{fig:region_based_intensities}
\end{figure}

The RSF model \cite{li2008minimization} modifies these constants into region-based intensities (Figure \ref{fig:region_based_intensities}):
\begin{equation}
\label{eqn:local_intensity}
r_\pm(\mathbf{x}) = \frac{K(\sigma_1, \mathbf{x}) \ast H_\pm\left[\phi(\mathbf{x})\right] I(\mathbf{x})}{{K(\sigma_1, \mathbf{x}) \ast H_\pm\left[\phi(\mathbf{x})\right]}}
\end{equation}
where $K$ is an $n$-dimensional Gaussian kernel:
\begin{equation}
\label{deqn_ex1a}
K(\sigma, \mathbf{u}) = \frac{1}{\sqrt[d]{2\pi\sigma^2}}\exp\left[{\frac{-\mathbf{u}^T\mathbf{u}}{2\sigma^2}}\right]
\end{equation}
that limits the intensity approximation to pixels in the neighborhood of $\mathbf{x}$. Similarly, a Heaviside function $H_\pm$ limits the contribution to pixels inside or outside of the contour, based on the sign:
\begin{align}
    H_+\left[\phi(\mathbf{x})\right] &= H_\epsilon\left[\phi(\mathbf{x})\right]\\
    H_-\left[\phi(\mathbf{x})\right] &= 1-H_\epsilon\left[\phi(\mathbf{x})\right]
\end{align}
A continuous approximation to the Heaviside function is used:
\begin{equation}
    H_\epsilon(u) = \frac{1}{2}\left[ 1 + \frac{2}{\pi}\arctan{\frac{u}{\epsilon}}\right]
\end{equation}
to avoid division by zero in the denominator of $r$. A small constant $\epsilon$ is used as a regularization term and set based on the numerical precision.

The forces applied in the outward (expanding) and inward (squeezing) directions of the contour are given by:
\begin{equation}
\label{eqn:rsf_directional_force}
    F_\pm(\mathbf{x}, r_\pm) = \oiiint_{\Omega^\pm} K(\sigma_2, \mathbf{x}-\mathbf{y}) \left| I(\mathbf{y}) - r_\pm(\mathbf{x}) \right|^2 d\mathbf{y}
\end{equation}
where $F_-$ applies force from the inside of $\Omega$ and $F_+$ applies force from the outside. The final $n$-dimensional force term is:
\begin{equation}
    F(\mathbf{x}, \phi, I)=-\left[F_+(\mathbf{x},r_+) - F_-(\mathbf{x}, r_-) \right]
\end{equation}

Incorporating the smoothing and regularization terms into the level set equation (Equation \ref{eqn:levelset}), the final energy function is:
\begin{equation}
    E(\phi, I)=\left[\nabla^2 \phi - \frac{\nabla^2 \phi}{|\nabla \phi|} \right] + \delta(\phi) \left( \alpha \left[ \frac{\nabla^2 \phi}{|\nabla \phi|} \right] + \beta F(\phi, I) \right)
\end{equation}
where the impulse function:
\begin{equation}
    \delta(u) = \frac{1}{\pi} \frac{\epsilon}{\epsilon^2 + u^2}
\end{equation}
is the derivative of $H_\epsilon$ and constrains the smoothing and force terms to the neighborhood of the contour.

\section{Implementation Details}

The proposed RSF formulation facilitates parallel computation since the information used to evolve the contour is constrained to the neighborhood of $\mathbf{x}$ by $K$. This paper leverages two forms of parallelism for large-scale image volumes: \textit{volume level} and \textit{voxel level}. Volume level parallelism breaks the data set into cubic sub-volumes for independent processing. This approach takes advantage of the structure of microvascular networks, where connectivity within the volume is high and interconnections between volumes are localized. Voxel level parallelism takes advantage of SIMD graphics processors (GPUs) by parallelizing the level set calculation and evolution independently at each grid point.

\subsection{Volume Level Parallelism}

The mesh-like structure of microvessels results in small and highly localized cross-sections when the network is cut by a plane. This provides advantages for parallel contour evolution, since oir formulation leverages template calculations for evaluating gradients, curvature, and the local region-based fitting term. Contour cross-sections that cross region boundaries are small and highly localized, making it unlikely for errors to propagate across the contour. In addition, sections are much more likely to occur \textit{across} vessels instead of \textit{along} them, resulting in contours that primarily propagate along the volume boundary where gradient information is more reliable. 

Our approach leverages these features by dividing large volumes into manageable $500^3$ sub-volumes processed independently on a GPU. Template artifacts are minimized using a curtain region scaled by $\max (\sigma_1,\sigma_2)$.

\begin{figure}[h]
    \centering
        \includegraphics[width=0.75\textwidth]{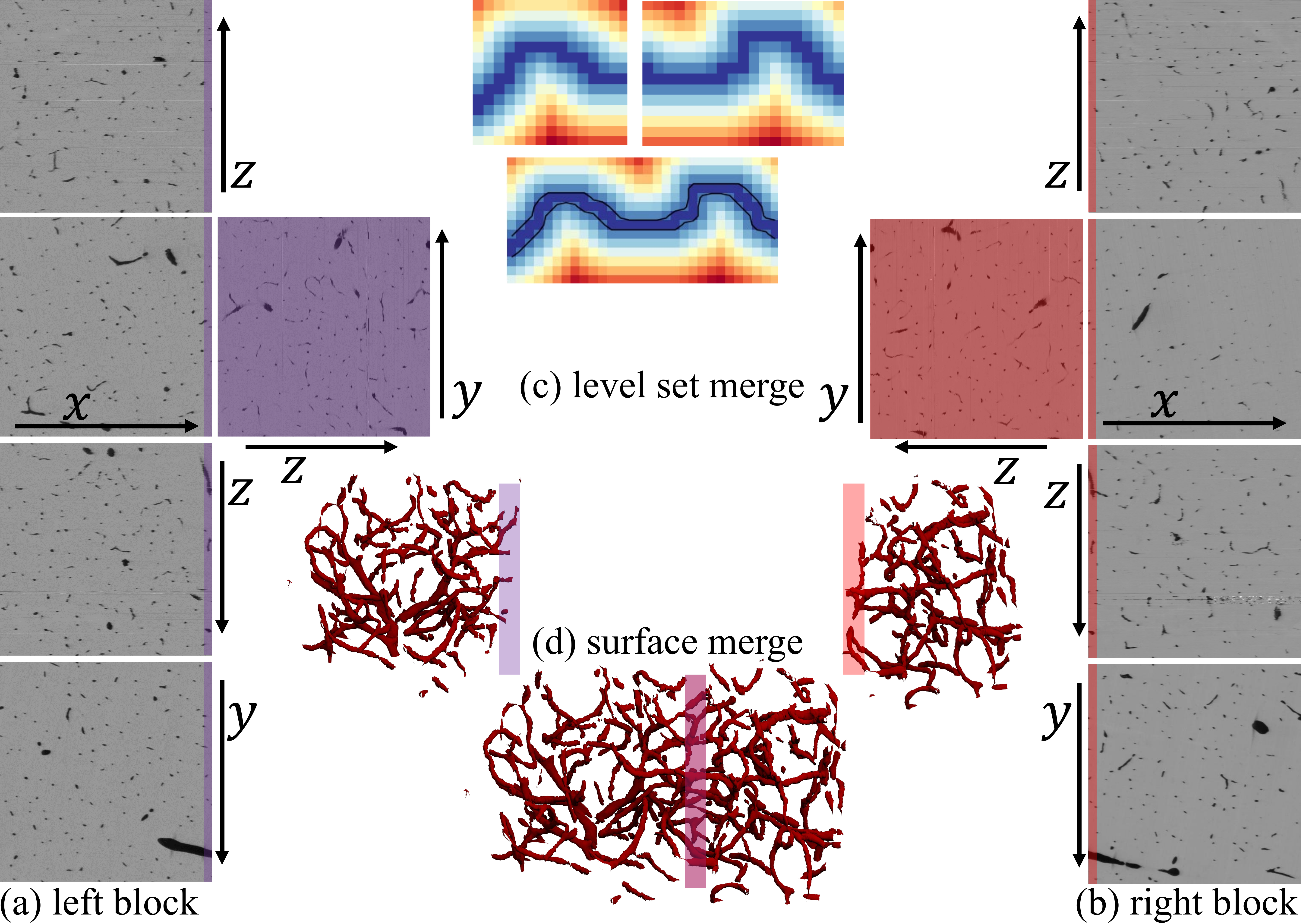}
        \caption{Linear interpolation is used to merge adjacent level set functions (a and b) $\phi_n$ (c) across overlapping curtain regions (purple/red). The merged $\phi_n$ is then used to generate the final contour (d).}
    \label{fig:merging_volumes}
\end{figure}

Neighboring volumes are merged using linear interpolation across the curtain region (Figure \ref{fig:merging_volumes}), which provides the best results when compared to the minimum, maximum, and averaging.

%We conducted experiments involving various merging techniques for the adjacent contours of $\phi_n$ to identify the optimal method for detecting our desired contour without altering its characteristics. As depicted in Figure \ref{fig:compare_merging}, it is evident that linear interpolation effectively detects the merging contour of $\phi$ without any distortion. Consequently, we adopted linear interpolation as the preferred method for merging our $500^3$ sub-volumes.
%\begin{figure}[tbh]
%    \centering
%        \includegraphics[width=\columnwidth]{figures/levelsetadd.pdf}
%        \caption{Comaping some merging techniques to adjacent contours (a)is performed by linear interpolation of $\phi_n$ (b)is performed by considering the minimum intensity of the overlapping region of $\phi_n$ (c)is performed by considering the maximum intensity of the overlapping region of $\phi_n$ (d)is performed by considering the average intensity of the overlapping region of $\phi_n$.}
%    \label{fig:compare_merging}
%\end{figure}

\subsection{Voxel Level Parallelism}

\begin{figure}[h]
    \centering
        \includegraphics[width=0.75\textwidth]{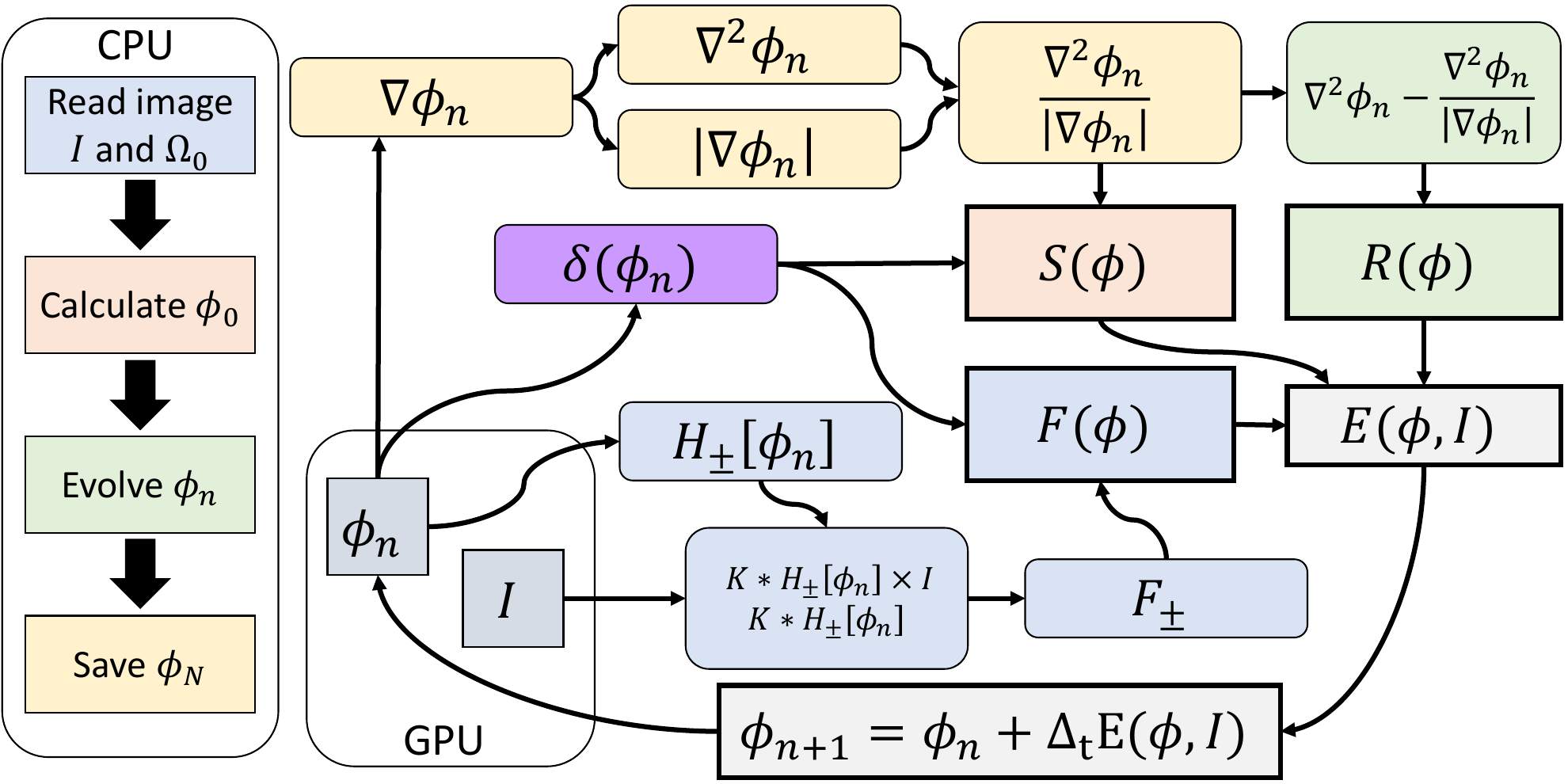}
        \caption{Outline of the proposed 3D RSF algorithm components implemented on the CPU and GPU. The Evolve component is performed completely using CUDA kernels to calculate $E$, with dependencies listed to minimize redundant computation.}
    \label{fig:schematic}
\end{figure}

The directional force (Equation \ref{eqn:rsf_directional_force}) requires two convolutions:
\begin{equation}
    F_\pm(\mathbf{x},r_\pm) = \left[ K*I^2 \right](\mathbf{x}) - 2r_\pm(\mathbf{x})\left[ K*I \right](\mathbf{x}) + r_\pm^2(\mathbf{x})
\end{equation}
where $K$ is a separable Gaussian kernel. We decompose the level set evolution terms into independently-evaluated kernels (Figure \ref{fig:schematic}) implemented using CUDA.

The source image $I(x,y,z)$ and the initial level set $\phi_0$ are transferred to the GPU, and all updates occur locally. Iterative updates to $\phi$ follow two major paths with kernel dependencies shown in Figure \ref{fig:schematic}. Kernels with all satisfied dependencies can be executed on available hardware as resources permit:
\begin{itemize}
    \item \textbf{$S(\phi)$ and $R(\phi)$ Path:} The gradient kernels are used to calculate the Laplacians for both the smoothing and regularization terms.
    \item \textbf{$F_{\pm}(\phi)$ Path:} The Heaviside functions and associated convolutions with the input image are used to calculate region-based intensities and force terms.
\end{itemize}
After the smoothing, regularization, and force terms are computed, they are combined into an energy term $E(\phi, I)$ used to update $\phi$.

\subsection{Initialization}
\label{sec:initialization}
In this section, we present a method for specifying the initial contour by calculating a set of seed points on the network. %\david{I commented most of this stuff out because it isn't really necessary. Be short and precise about what you're doing. Specify the algorithm and how its used. For example, "We set an initial threshold using Otsu's method and then identify edges using a Canny edge detector."}

\begin{figure}[h]
    \centering
%    \begin{minipage}{2.5in}
        \includegraphics[width=0.75\textwidth]{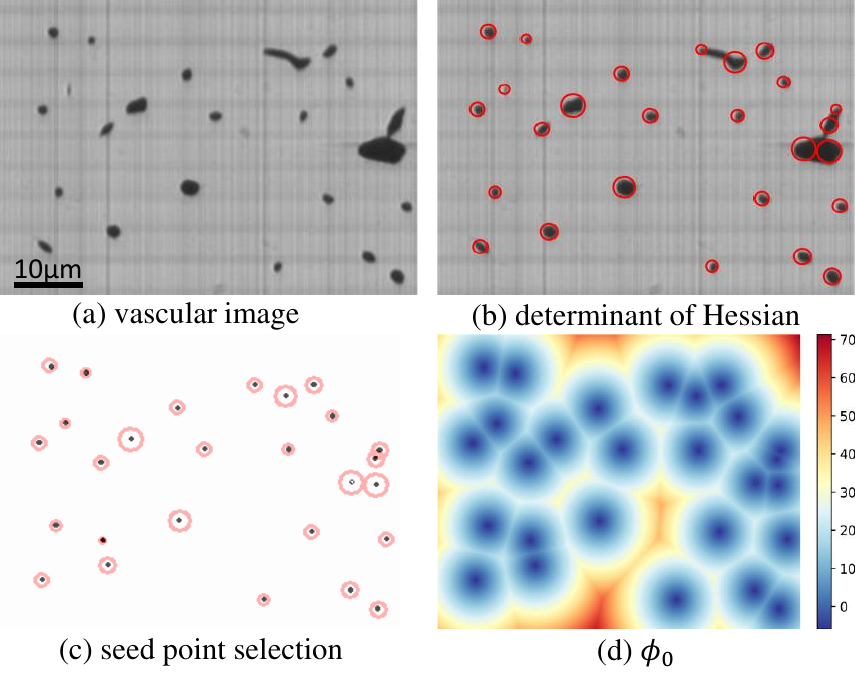}
        \caption{Initializing the level set from an input image (a). Blob detection using the determinant of the Hessian (b) is used to identify candidate seed points (c) on the vascular centerlines. Fast sweeping is then used to determine the positive distance (d) from seed points to create the $\phi_0$.}
%    \end{minipage}
    \label{fig:seed_points}
\end{figure}

%We set an initial thresholding using a vesselness filter \cite{frangi1998multiscale} to convert the input image into a binary representation. Next, we use a Canny edge detector to enhance vessel edges and thus distinguish vessels from the background. 
The determinant of the Hessian matrix blob detection \cite{xu2014blob} is used to identify a set of candidate seed points $\mathbf{S}$ in each $z$-axis slice. The level set $\phi_0$ is initialized by setting these seed point positions to zero: $\phi_0(\mathbf{x})=0$ if $\mathbf{x}\in \mathbf{S}$. $\phi_0$ is then initialized to a signed distance function using the fast sweeping algorithm \cite{zhao2005fast}.

\section{Results}
We first demonstrate the proposed RSF formulation on phantom 2D images with varying noise and contrast levels. We then apply the proposed 3D implementation to several types of 3D microscopy capable of imaging large tissue volumes.

\subsection{2D Validation}
\begin{figure}[h]
    \centering
        \includegraphics[width=0.82\textwidth]{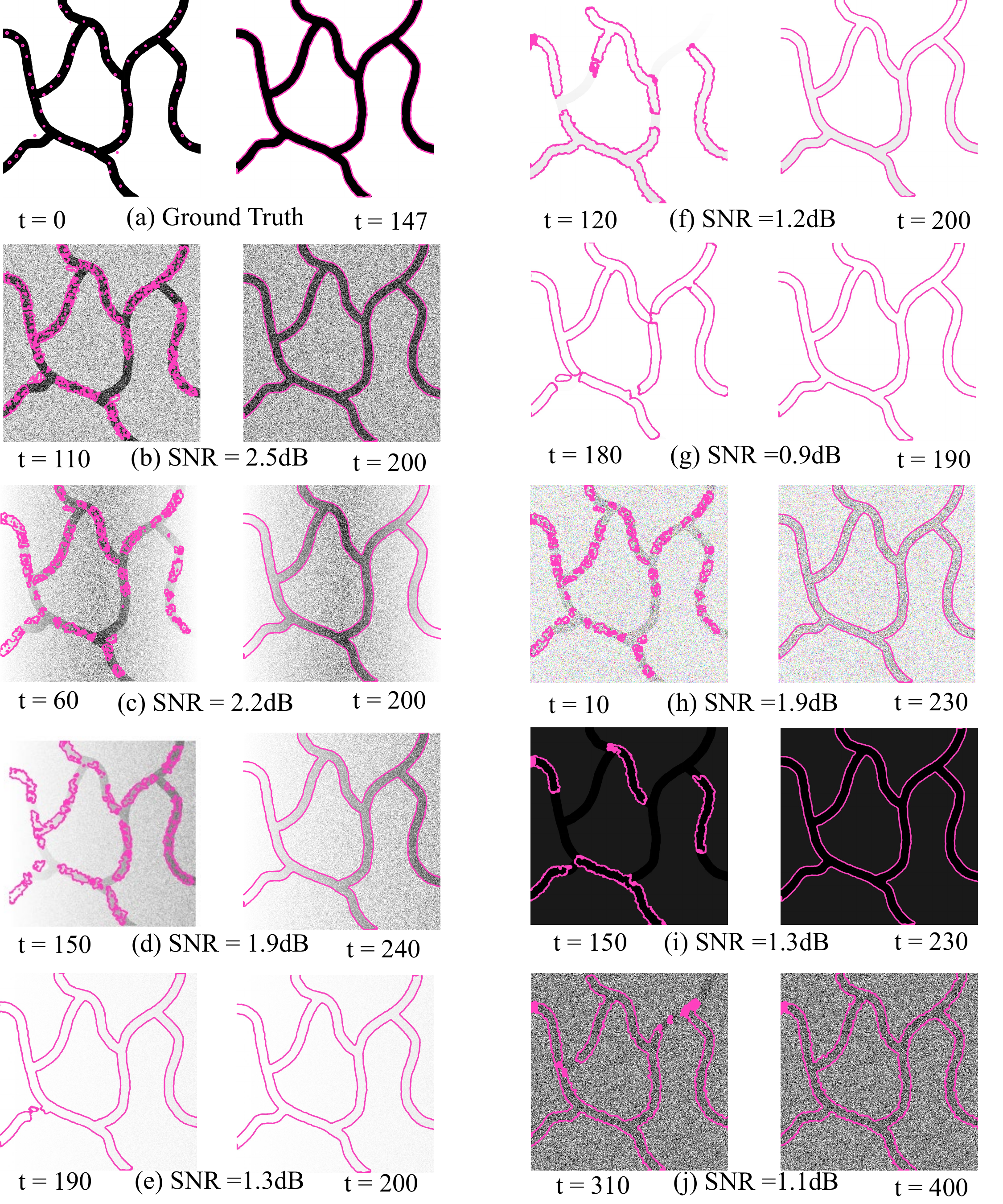}
        \caption{Results of RSF method for 2D synthetic vasculature. The curve evolution process from the initial contour (in the first and third columns) to the final contour (in the second and fourth columns) is shown in every row for the corresponding image. We altered the Signal-to-Noise Ratio (SNR) of the ground truth in certain images through two distinct methods. Firstly, we introduced Gaussian noise, and secondly, we modified the intensity along the x-axis to the y-axis for some images and vice versa along the y-axis to the x-axis for others.}
        \label{fig:2D_validation}
\end{figure}

We first evaluate the performance of the proposed model using 2D synthetic images. While 2D performance has been previously reported \cite{wang2021review}, our images exhibit interconnected topological features comparable to microvascular networks with contrast and noise variations (Figure \ref{fig:2D_validation}). The raw image is shown with seed points and the final surface (Figure \ref{fig:2D_validation}a). Additional images (Figure \ref{fig:2D_validation}b - \ref{fig:2D_validation}j) show the level set evolution at intermediate and final time steps for images with various noise and contrast artifacts. Parameter values for all these segmentations were as follows: $\sigma_1$ = 19, $\sigma_2$ = 9, $\beta = 3.5$, $\alpha$ = $255\times255\times0.01$ and $dt=0.1$.

\begin{figure}[tbh]
    \centering
        \includegraphics[width=0.75\textwidth]{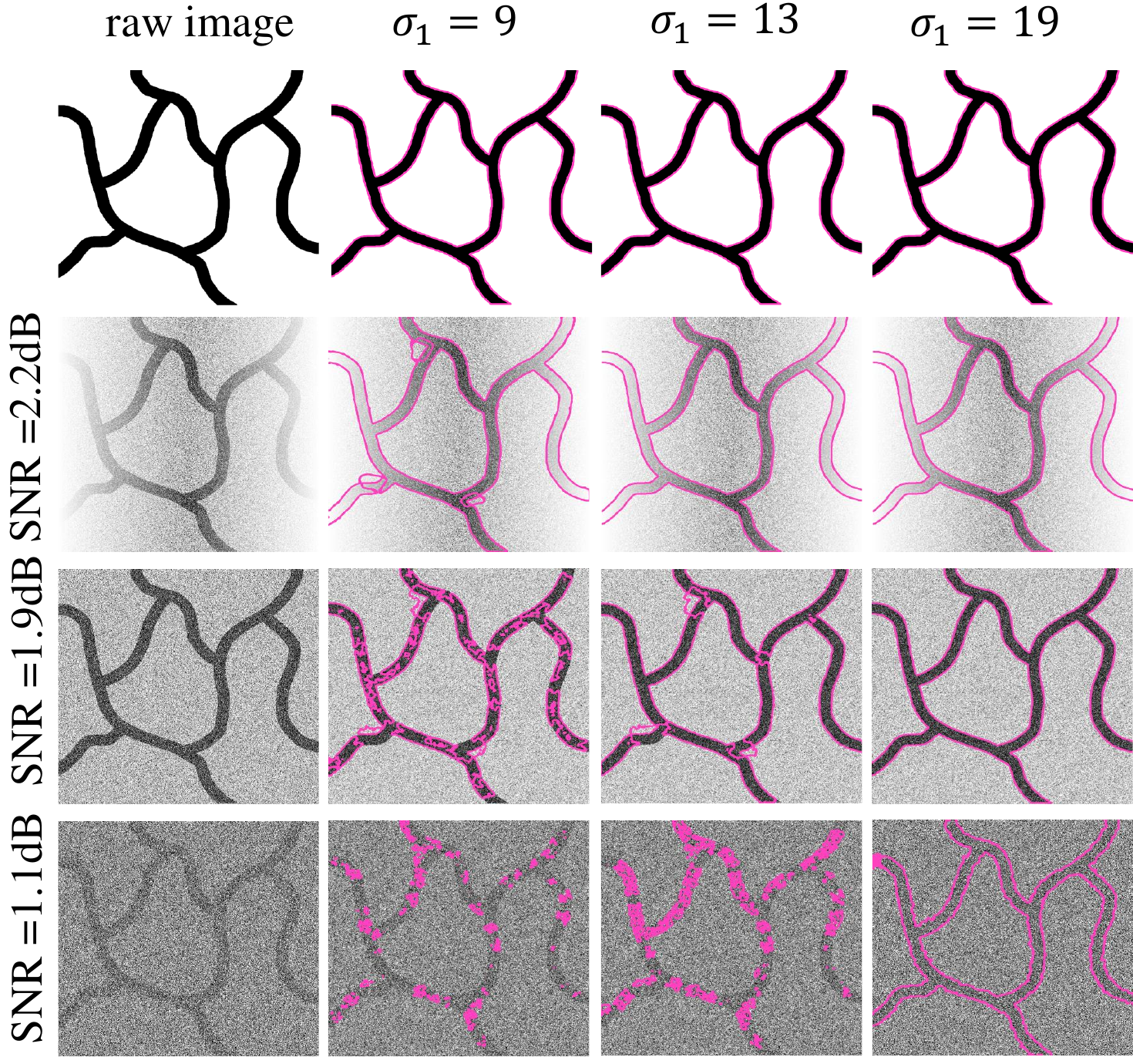}
        \caption{Results of 2D synthetic images for various $\sigma_1$ values used to calculate region-based intensities. This demonstrates that the algorithm is insensitive to high $\sigma_1$ values for these images.}
    \label{fig:sigma_change}
\end{figure}

The parameter most sensitive to image noise is $\sigma_1$ used to calculate the region-based intensities. However, after testing various combinations, we found that higher values (ex. $\sigma_1 = 19$) were robust for all cases (Figure \ref{fig:sigma_change}).

\subsection{Vascular Image Acquisition}
\begin{figure}[ht]
    \centering
        \includegraphics[width=0.65\textwidth]{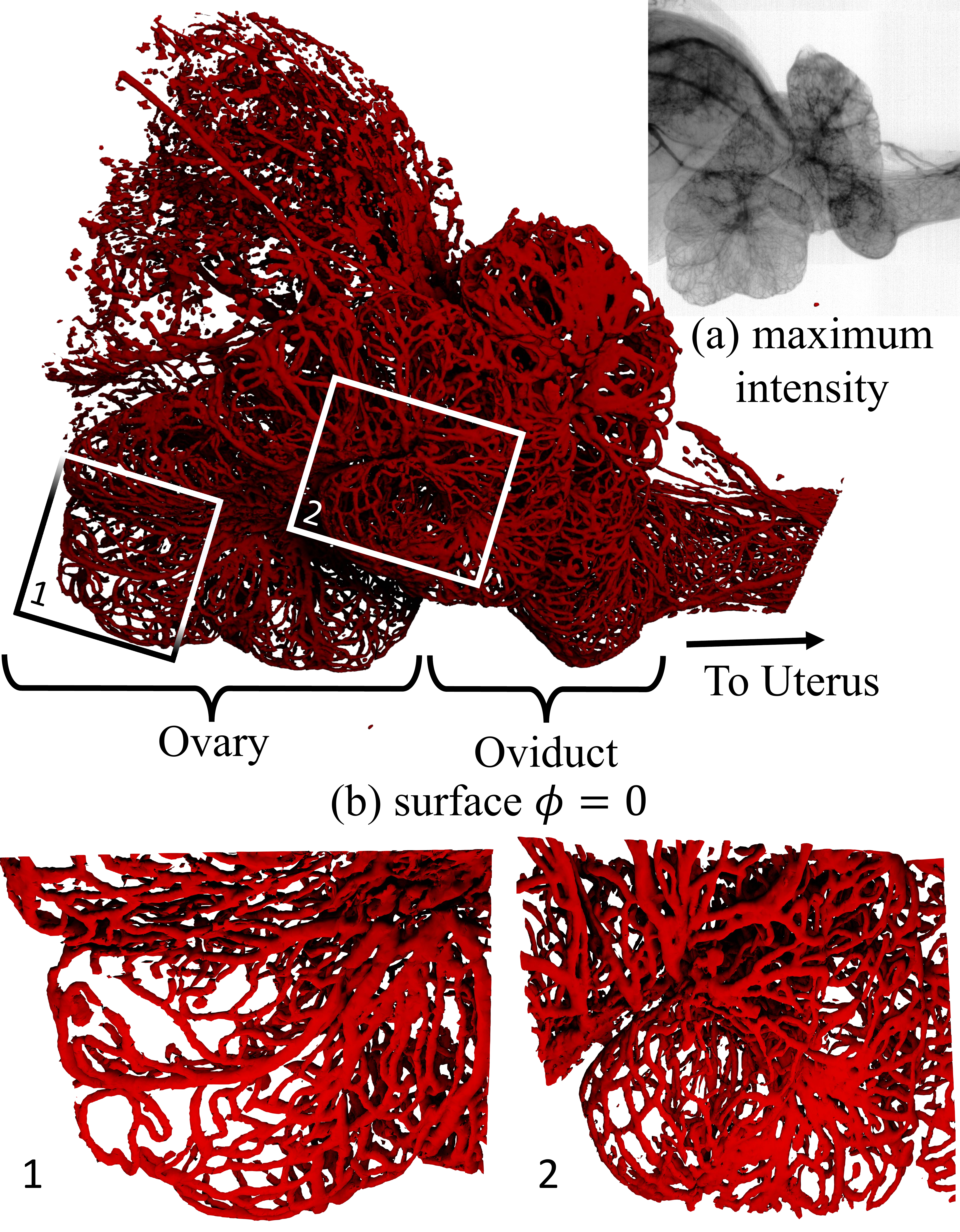}
        \caption{Ovary from a female mouse (\textit{BRaf\textsuperscript{V600E/WT}}) at age 2 months and 23 days. Imaging modality: LSFM, Label: lectin-649, Resolution:  \(0.949\,\mu\text{m} \times 0.949\,\mu\text{m} \times 6\,\mu\text{m}\), Data size: $6.6\,\text{GB}$. (a) The maximum intensity projection is shown along with (b) the vascular surface where $\phi=0$. Two labeled insets show magnified regions of the ovary.}
        \label{fig:ovary}
\end{figure}

\begin{figure*}[h]
    \centering
        \includegraphics[width=\textwidth]{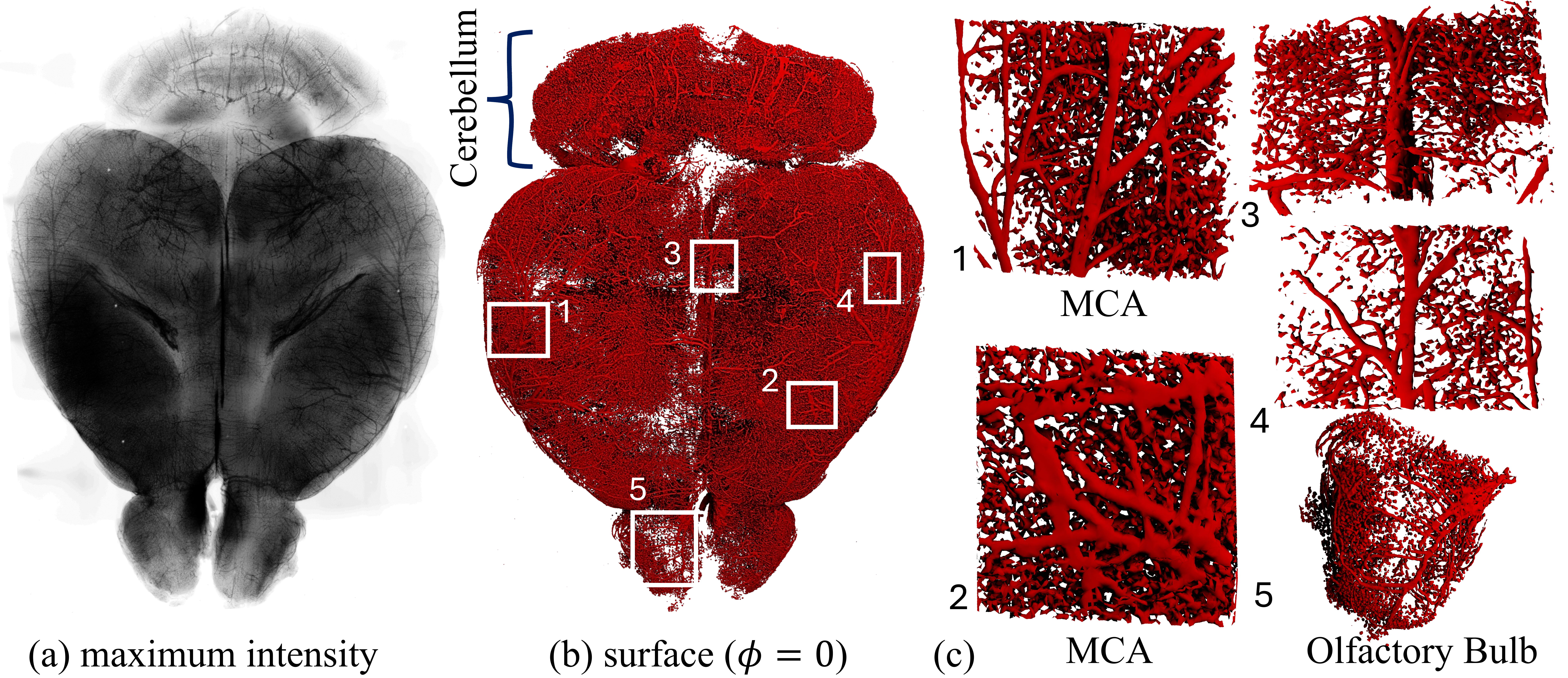}
        \caption{Light sheet microscopy image of a 4.5-month-old female C57BL/6N mouse brain. The vasculature is labeled with a \SI{488}{\nano\meter} fluorescently conjugated tomato lectin. The spatial resolution is \SI{2}{\micro\meter} $\times$ \SI{2}{\micro\meter} $\times$ \SI{6}{\micro\meter} with a total data size of 103.6 GB. (a) The maximum intensity projection of the whole brain is shown on the left, with (b) the complete reconstructed network. Cropped magnified regions are shown as magnified insets (c). 1 and 2 show the right and left middle cerebral artery (MCA), and 5 shows the olfactory bulb.}
        \label{fig:Brain_lightsheet}
\end{figure*}

\begin{figure*}[ht]
    \centering
    \includegraphics[width=\textwidth]{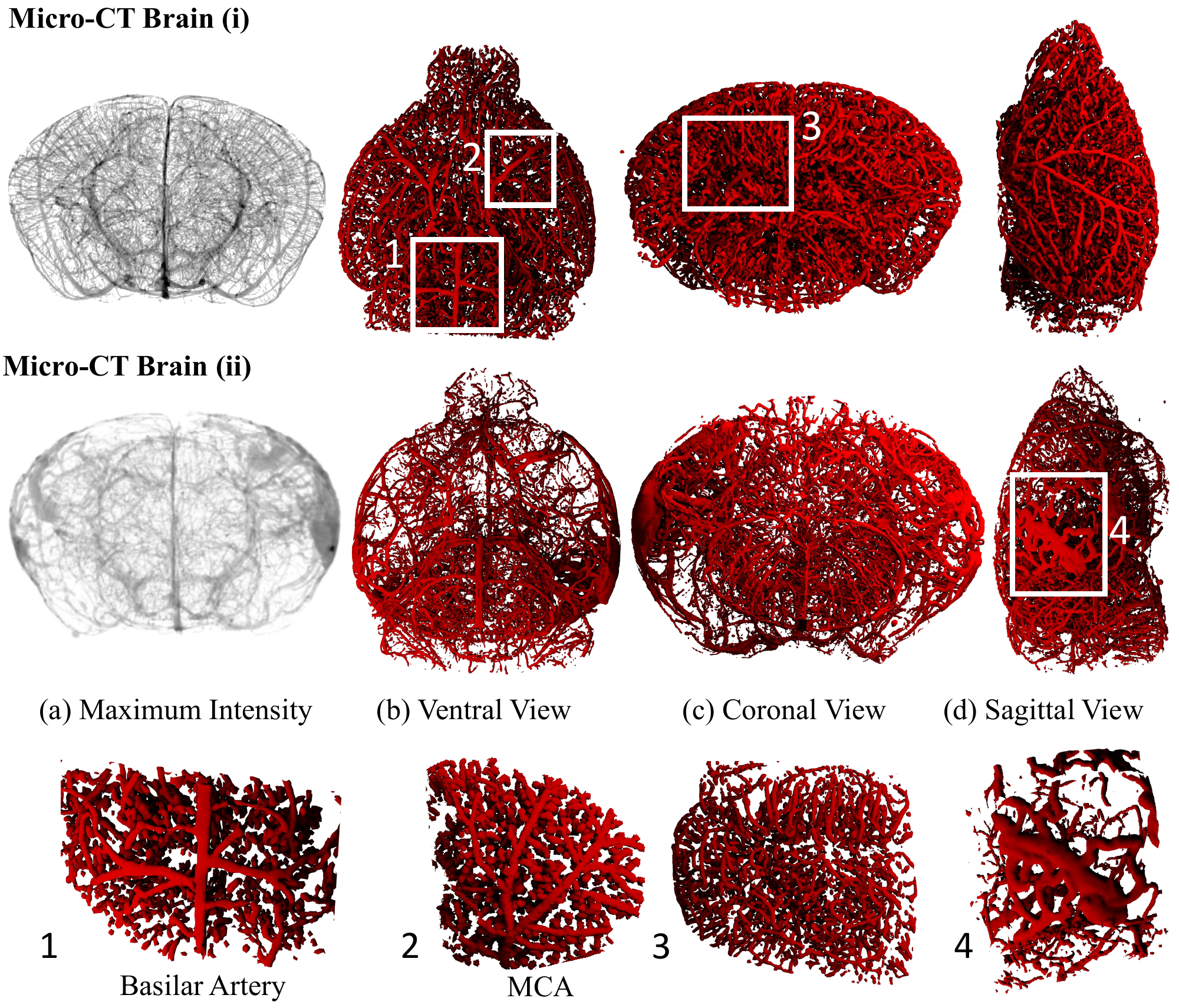}
        \caption{Two 2-month-old female mouse brains (i) and (ii) with
        the vasculature perfused with Vascupaint and imaged using a Skyscan 127. The spatial resolution image is \SI{10}{\micro\meter} $\times$ \SI{10}{\micro\meter} $\times$ \SI{10}{\micro\meter}. (a) Maximum intensity projections are shown along side a (b) ventral view, (c) coronal view, and (d) sagittal view of the reconstructed networks. Cropped magnified regions of these two networks are shown as insets (1-4).}
        \label{fig:Brain_micro-ct}
\end{figure*}

\begin{figure*}[h]
    \centering
        \includegraphics[width=\textwidth]{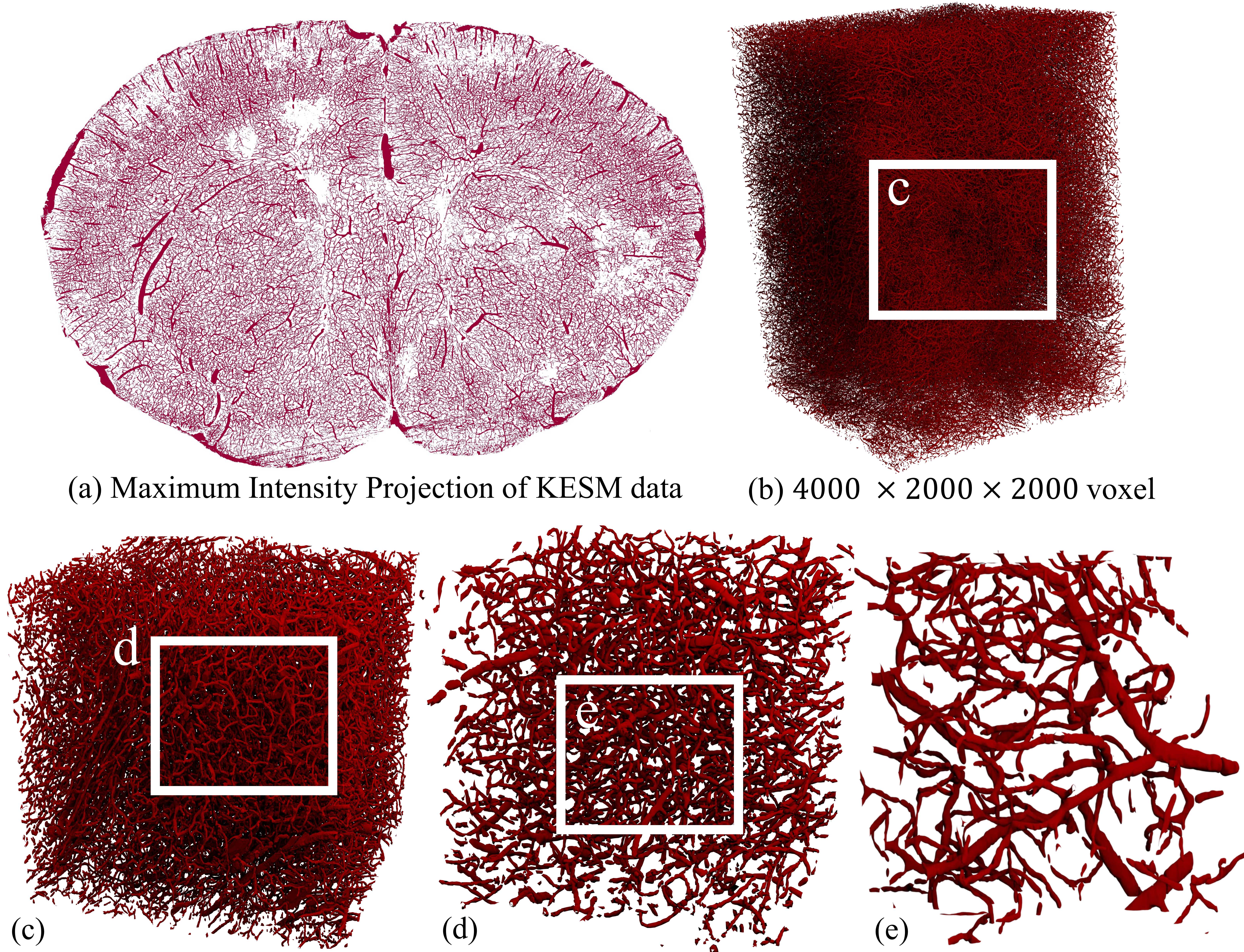}
        \caption{Whole brain vasculature imaged using knife edge scanning microscopy (KESM) and available online (kesm.cs.tamu.edu)\cite{chung2011multiscale} and imaged at a resolution of \SI{0.6}{\micro\meter}$\times$\SI{0.7}{\micro\meter}$\times$\SI{1}{\micro\meter} \cite{mayerich2011fast}. (a) Reconstruction from 140 slices across the whole brain is shown, along with (b) a 4000$\times$2000$\times$2000 voxel sub-volume with (c-e) several iterated higher-resolution zooms. }
        \label{fig:large_scale}
\end{figure*}

Five comprehensive 3D images of tissue microvasculature were collected using micro-CT (2 brains), light sheet microscopy (brain and ovary), and knife-edge scanning microscopy (brain):
\begin{itemize}
     \item \textbf{Mouse ovary, LSM} (Figure \ref{fig:ovary}): An ovary from a 2-month and 23 days-old female mouse  (\textit{BRaf\textsuperscript{V600E/WT}}; \textit{Ai14 R26\textsuperscript{-lsl-TdTom/+}} ) perfused with lectin-649 (Vector Laboratories, Product number: DL-1178-1; Newark, CA, USA) and imaged using a Zeiss Z7 (Zeiss, Oberkochen, Baden-Württemberg, Germany) light sheet microscope. The voxel size of the image is \(0.949\,\mu\text{m} \times 0.949\,\mu\text{m} \times 6\,\mu\text{m}\), and the file size is $6.6\,\text{GB}$. This image was obtained with 5X lens, 638 nm laser, 18\% laser power, 249.67 ms exposure time.
    \item \textbf{Mouse brain, LSM} (Figure \ref{fig:Brain_lightsheet}): A brain from a 4.5-month-old female mouse (C57BL/6N) was perfused with lectin-488 (Vector Laboratories, Product number: DL-1174-1; Newark, CA, USA) and imaged by AxL Cleared Tissue LightSheet (Intelligent Imaging, Denver, CO, USA). The voxel size of the image is \(2\,\mu\text{m} \times 2\,\mu\text{m} \times 6\,\mu\text{m}\), and the file size is $103.6\,\text{GB}$. This image was acquired with a 1X lens, 488 nm laser, 200 mW laser power, and 300 ms exposure time.
    \item \textbf{Mouse brain, $\mu$-CT} (Figure \ref{fig:Brain_micro-ct}) Two separate 2 month-old female mouse brains (\textit{Tg(Slco1c1-BAC-CreER)}; \textit{R26\textsuperscript{-lsl-TdTom/+}}) were perfused with Vascupaint (MediLumine, Product number: MDL-121; Montreal, Quebec, Canada) and imaged with a Skyscan 1276 (Bruker, Billerica, MA, USA). The imaging was performed at an isotropic voxel size of \SI{10}{\micro\meter}.
    \item \textbf{Mouse brain, KESM} (Figure \ref{fig:large_scale}): Whole mouse mouse brain (C57BL/6J) was perfused with India ink \cite{mayerich2011fast} and imaged at a resolution of \SI{0.6}{\micro\meter} $\times$ \SI{0.7}{\micro\meter} $\times$ \SI{1}{\micro\meter}.
\end{itemize}
 
%\subsection{Microvascular Network Reconstruction}

%\textcolor{blue}{We segmented a $4000 \times 2000 \times 2000$ voxel volume of mouse brain microvascular imagery collected using knife-edge scanning microscopy (KESM). The full volume is shown (Figure \ref{fig:large_scale}a) along with progressively higher-magnification visualizations to convey connectivity and topology (Figure \ref{fig:large_scale}b - d).}

%\textcolor{blue}{To enhance the visual representation and provide a more detailed perspective, in Figure \ref{fig:large_scale} we created Volumetric projection and then illustrations of the $4000 \times 2000\times2000$ voxel dataset. We then zoomed in on specific regions and presented them in columns 2, 3, and 4, respectively. This approach allows for a closer examination of specific areas within the larger dataset, providing a comprehensive view of the finer details. }

\subsection{3D Monte-Carlo Validation}
The size and complexity of microvascular networks make extensive manual annotation impractical. To test the viability of the proposed RSF-based level set approach as a segmentation method, we employed Monte-Carlo-based validation.

\begin{figure*}[h]
    \centering
    \includegraphics[width=\textwidth]{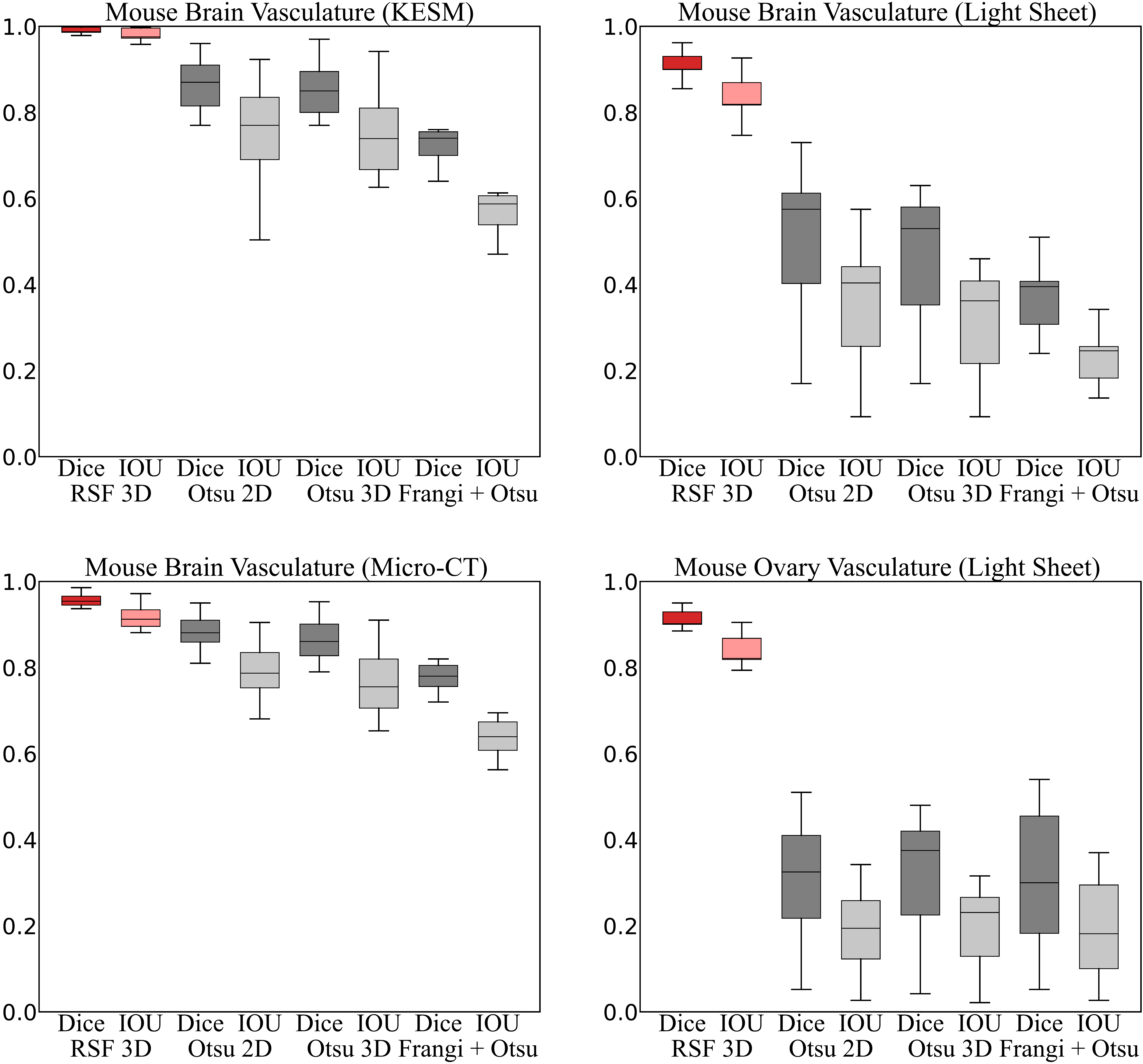}
    \caption{Segmentation results for mouse microvascular networks across modalities. Results are expressed using both the Dice coefficient (dark) and Jaccard index (light). The GPU-based 3D RSF approach proposed here is shown in red. Comparisons use Otsu's method with and without pre-processing with a vesselness filter \cite{frangi1998multiscale}.}
    \label{fig:3D_validation}
\end{figure*}

A randomized set of $100\times 100 \times 100$ voxel sub-volumes were randomly selected from each image: 13 from KESM, 12 from both CT data sets, and 6 from each LSFM data set. These volumes were manually annotated using 3D Slicer \cite{pieper20043d} and segmented with the proposed RSF method. Statistics for the Dice coefficient \cite{duarte1999comparison} and Jaccard index \cite{real1996probabilistic} were calculated (Figure \ref{fig:3D_validation}). All our 3D segmentations used the same parameters for level set evolution: $\sigma_1$ = 5, $\sigma_2$ = 0, $dt = 0.06$, $\alpha = 255\times255\times0.0009$, $\beta = 0.1$.

\begin{figure*}[h]
    \centering
        \includegraphics[width=\textwidth]{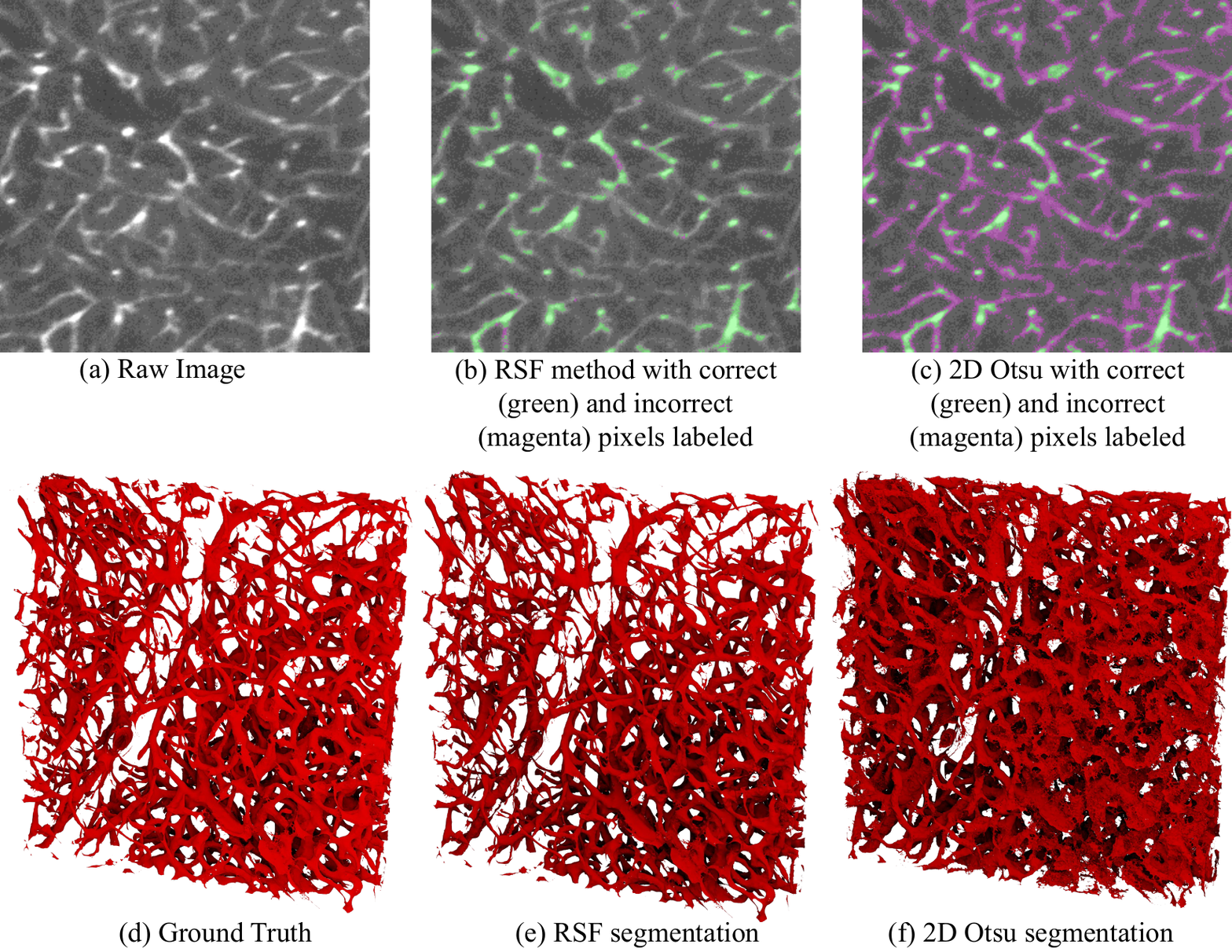}
        \caption{Segmentation results for RSF and Otsu's method using a 2D "slice-by-slice" threshold. (a) The raw image is shown with overlays showing correctly labeled pixels (green) with oversegmentation (magenta) using both (b) RSF and (c) Otsu. While the magenta pixels correspond to vessels, they are blurred fluorescence from adjacent slices caused by anisotropic sampling along the z-axis. This results in over-segmentation along the z-axis visible in a 3D visualization of the segmentation results (d-f).}
        \label{fig:comparison_visualization}
\end{figure*}

RSF results for the LSFM brain image are shown compared to the best-performing alternative (Figure \ref{fig:comparison_visualization}), which was Otsu's method applied to each z-slice section independently. Note that the primary problem is over-segmentation, particularly across the z-axis where the resolution is lowest. This results in the cumulative integration of errors across several sections, as seen in the model visualization (Figure \ref{fig:comparison_visualization}(f)).

\subsection{Performance Profiling}
In this section, we present a thorough analysis of the computational performance, focusing on a comparative study between GPU and CPU acceleration. The summarized results of this analysis can be found in Table \ref{table:profiling}, specifically addressing voxel-level parallelism. Our primary objective is to maximize GPU utilization in order to achieve peak performance. Notably, our GPU model exhibits a significant speedup compared to its CPU counterpart. We have successfully maintained consistent performance across all kernels, aligning with our overarching goal.

%Additionally, in Table \ref{table:whole_profiling}, we extend our profiling evaluation to encompass the entire volume-level parallelism. This assessment involves a volumetric dataset with dimensions $4000 \times 2000 \times 2000$, representing a substantial $16GB$ dataset.

%\begin{figure}[ht]
%    \centering
%        \includegraphics[width=\columnwidth]{figures/pie_chart.png}
%        \caption{GPU Acceleration Analysis for a 500x500x500 Cube: Timings, Parameter Breakdown,and CPU Speedup}
%\end{figure}
\begin{table}[ht]
    \centering
    \begin{tabular}{|>{\centering\arraybackslash}m{2cm}|>{\centering\arraybackslash}m{2cm}|>{\centering\arraybackslash}m{1.5cm}|>{\centering\arraybackslash}m{2cm}|>{\centering\arraybackslash}m{2cm}|}
    \hline
        Parameters & GPU & Percent & Speedup   \\ \hline
        $H_-I$ & 41 ms & 7.0\% & 92.6x  \\ \hline
        $H_+I$ & 41 ms & 7.0\% & 92.6x    \\ \hline
        $K*H_-I$ & 58 ms & 10\% & 281x  \\ \hline
        $K*H_+I$ & 57 ms & 9.7\% & 287x  \\ \hline
        $K*H_-$ & 56 ms & 9.6\% & 291x  \\ \hline
        $K*H_+$ & 58 ms & 10\% & 282x  \\ \hline
        $\delta(\nu)$ & 20 ms & 3.4\% & 165x\\ \hline
        $\nabla\phi$ & 50 ms & 8.5\% & 360x \\ \hline
        $|\nabla\phi|$ & 15 ms & 2.6\% & 386x  \\ \hline
        $\nabla^2\phi$ & 53 ms & 9.1\% & 358x  \\ \hline
        $\frac{\nabla\phi}{|\nabla\phi|}$ & 47 ms & 8.0\% & 102x  \\ \hline
        $\nabla^2\phi-\frac{\nabla\phi}{|\nabla\phi|}$ & 13 ms & 2.2\% & 192x \\ \hline
        $E_+$ & 38 ms & 6.5\% & 224x  \\ \hline
        $E_-$ & 38 ms & 6.5\% & 224x  \\ \hline
    \end{tabular}
    \caption{Profiling on GPU and CPU for a $500\times500\times500$ voxel.The first column represents the main parameters used in the RSF method. The time required for each time step in GPU is in the second column and in CPU is in the third column. In the fourth column we showed the speeding for each parameter from CPU to GPU.}
    \label{table:profiling}
\end{table}

\section{Conclusion}
We propose a reformulation of the RSF region-based level set method that is amenable to 3D images and parallel implementation for large data sets. We also provide a GPU-based implementation of this algorithm, including rigorous optimization and profiling for SIMD processors. 

While the efficacy of RSF has been demonstrated previously on more traditional biomedical images, this is the first demonstration of it on topologically complex structures from emerging microscopy imaging methods. The proposed approach bypasses long computation times and memory requirements by formulating an RSF algorithm that can be applied independently to subvolumes of the whole image and evolved entirely on the GPU.

We also demonstrate a validation method using Monte-Carlo sampling to assess the efficacy of 3D RSF segmentation using the Dice coefficient and Jaccard index. The graphical representations of these comparisons provide compelling evidence of its robustness to noise and contrast changes, as well as its applicability to topologically complex structures.

The significance of this work extends beyond the specific implementation of the RSF model by offering a versatile tool for accurate and efficient 3D microvascular segmentation. The integration of GPU acceleration not only improves computational efficiency but also opens new possibilities for large-scale image processing extending to whole organs.

\section*{Acknowledgements}
This work was supported by grants from the National Institutes of Health to JDW (1R01HL159159) and to JDW and DM (R01HL146745), the National Science Foundation to DM (1943455), as well as generous seed funding from the University of Virginia School of Medicine and UVA Comprehensive Cancer Center to JDW.

DM is a stakeholder in SwiftFront, LLC.

 \bibliographystyle{elsarticle-num} 
 \bibliography{vesselseg}





% \end{thebibliography}
\end{document}